# Pb(Mg$_{1/3}$Nb$_{2/3}$)O$_3$ (PMN) relaxor: dipole glass or nano-domain ferroelectric ?

Desheng Fu[1], Hiroki Taniguchi[2], Mitsuru Itoh[2], and Shigeo Mori[3]

[1]*Division of Global Research Leaders, Shizuoka University, Johoku 3-5-1, Naka-ku, Hamamatsu 432-8561, Japan*
*E-mail:ddsfu@ipc.shizuoka.ac.jp*
[2]*Materials and Structures Laboratory, Tokyo Institute of Technology, 4259 Nagatsuta, Yokohama 226-8503, Japan*
[3]*Department of Materials Science, Osaka Prefecture University, Sakai, Osaka 599-8531, Japan*

## Abstract

Combining our comprehensive investigations of polarization evolution, soft-mode by Raman scattering and microstructure by TEM, and the results reported in the literatures, we show that prototypical relaxor Pb(Mg$_{1/3}$Nb$_{2/3}$)O$_3$ (PMN) is essentially ferroelectric for $T<Tc$~225 K. Its anomalous dielectric behavior in a broad temperature range results from the reorientation of domains in the crystal. A physic picture of the structure evolution in relaxor is also revealed. It is found that nanometric ferroelectric domains (gennerally called as polar nano-region (PNR)) interact cooperatively to form micrometric domain. Such multiscale inhomogeneities of domain structure in addition to the well-known inhomogeneities of chemical composition and local symmetry are considered to play a crucial role in producing the enigmatic phenomena in relaxor system.

## 1. Introduction

In sharp contrast to normal ferroelectric (for example BaTiO$_3$), relaxors show unusually large dielectric constant within a large temperature range (~100 K) (Fig. 1 (a)) [1,2]. Such dielectric response is strongly dependent on the frequency. Its origin has been the focus of interest in the solid-state physics. Unlike the dielectric anomaly in BaTiO$_3$, which is associated with a ferroelectric phase transition, the maximum of dielectric response doesn't indicate the occurrence of ferroelectric phase transition. Such huge dielectric response suggests that local polarization might occur in the crystal. This was envisioned by Burns and Dacol from the deviation from linearity of refractive index $n(T)$ (Fig. 1(c)) around the so-called Burns temperature ($T_{Burns}$) [3] because the deviation $\Delta n$ is proportional to polarization $P_s$. The local polarization is suggested to occur in a nano-region, and is generally called as polar nano-region (PNR). The existence of PNR in relaxor is well confirmed from the neutron scattering measurements [4] and transition electron microscopy (TEM) observation. However, it is still unknown how PNRs contribute to the large dielectric response. More recently, it is also suggested that a strong coupling between zone-center and zone-boundary soft-modes may play a key role in understanding the relaxor behaviors [5]. Clearly, the question "what is the origin of the giant dielectric constant over a broad temperature range?" is still unclear [6,7].

Another longstanding issue on relaxors is how PNRs interact at low temperature. There are two acceptable models: (1) dipole glass model [2, 8-11], and (2) random-field model [12,13]. In spherical random-bond–random-field (SRBRF) model, Pirc and Blinc assumed that PNRs are spherical and interact randomly and proposed a frozen dipole glass state for relaxor (Fig. 2(a)) [10,11]. It predicts that the scaled third-order nonlinear susceptibility $a_3 = -\varepsilon_3/\varepsilon_1^4$ will shows a nearly divergent behavior at the freezing temperature $T_f$ of the spherical glass phase. In sharp contrast, the random field model of Fisch [13], which assumes **non**-random two-spin exchange, predicts a ferroelectric or ferroelectric domain state in relaxor (Fig. 2(b)). This random Potts field model also predicts a broadening specific heat peak for the glass phase and shows that the latent heat at ferroelectric transition $T_c$ is so small that it may be difficult to be detected, which reasonably explains the data reported by Moriya et al. [14].

Combining our recent results [15] from polarization, Raman scattering, TEM measurements and those reported in the literature, here, we propose a physical picture to understand the behaviors of Pb(Mg$_{1/3}$Nb$_{2/3}$)O$_3$ (PMN) relaxor.

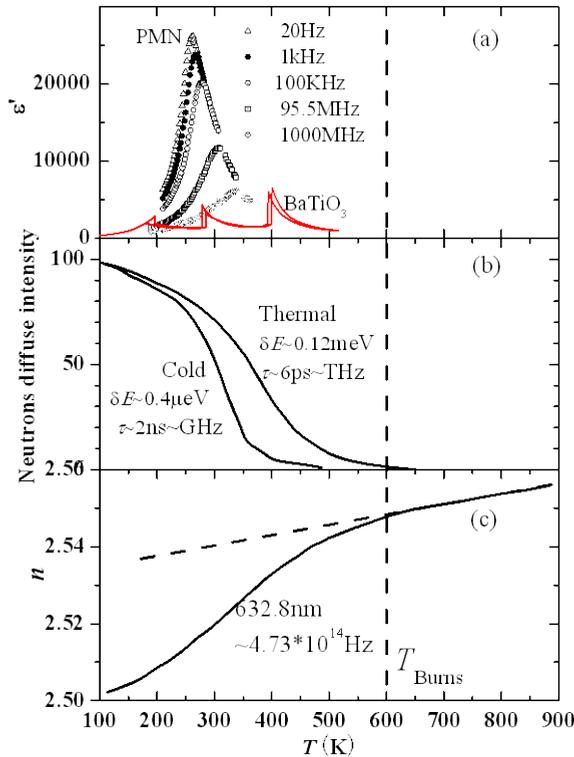

**Fig. 1.** Temperature dependence of (a) linear dielectric constant [9], (b) cold [35] & thermal [4] neutrons diffuse intensities for PMN relaxor crystal, and (c) refractive index [3].

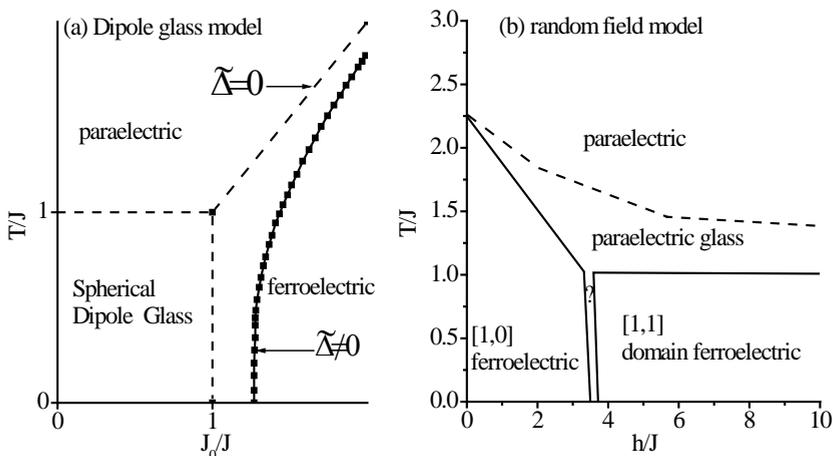

**Fig. 2.** Two phase diagrams proposed for relaxor by (a) Dipole glass model [10], and (b) Random field model [13].

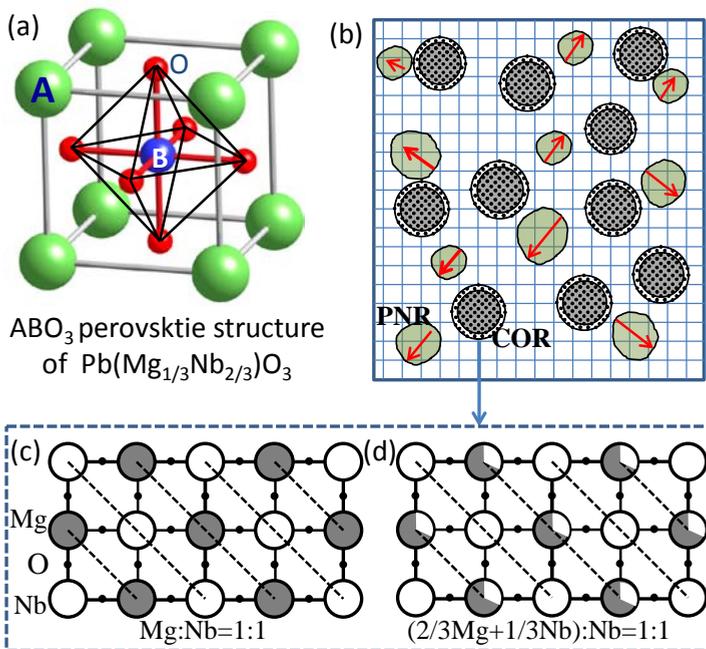

**Fig. 3.** (a) ABO$_3$ perovskite structure. (b) Model for relaxor structure. PNR and COR represent the polar nano-region and chemically order region, respectively. (c) & (d) show two models of atom arrangement for COR. To maintain the electric neutrality, a Nb-rich layer is required for case (c).

## 2. Multiple inhomogeneities in relaxors

PMN is a prototypical relaxor with A(B′,B″)$O_3$ perovskite structure (Fig. 3), in which B-site is occupied by two kinds of heterovalent cations. Such chemical inhomogeneity is a common feature of relaxor crystals. Although it remains an average centrosymmetric cubic structure down to 5 K [16], local structural inhomogeneity has been detected in PMN relaxor. In addition to PNR mentioned above, chemically ordering region (COR) [17-19] with size of several nm has been observed in PMN crystal by TEM. It should be noticed that PNR and COR belong to different symmetry groups and are considered to have non-centrosymmetric symmetry $R3m$ and centrosymmetric symmetry $Fm\bar{3}m$, respectively. Therefore, there is spontaneous polarization $P_s$ along the <111>$_c$ direction of pseudocubic structure in PNRs [20, 21], but is none of $P_s$ in CORs. In addition to chemical and structural inhomogeneities, we will show that PMN relaxor also has inhomogeneity of ferroelectric domain structure. Multiple inhomogeneities are thus considered to play a crucial role in inducing the intriguing relaxor behaviors.

## 3. Evolution of polarization and origin of huge dielectric responses

In order to understand the nature of huge dielectric response and ground state of polarization in PMN, it is essential to know the polarization hysteresis of all states including virgin state in PMN crystal. Although there are many reports on the polarization hysteresis of PMN crystal, there is a lack of understanding of the polarization hysteresis of the virgin state. In our polarization measurements, in order to access the virgin state of the crystal at a temperature, it was firstly annealed at 360 K and then cooled to the desired temperature for the measurement.

Figure 4 shows the $D$-$E$ hysteresis for three typical temperatures observed for (110)$_c$-cut PMN crystal. At $T$=360 K that is greatly higher than the freezing temperature $T_f$=224 K assumed for PMN crystal [9], there is no remnant polarization within the experimental time scale of $\tau$~10 ms (one cycle of the $D$–$E$ loop) and the polarization is history-independent, indicating that the crystal is macroscopically paraelectric at this temperature. When temperature is lower than room temperature (for example, $T$=250 K), remnant polarization was observed but it generally disappears after removing the electric field.

Upon further cooling to temperatures lower than ~220 K, PMN shows polarization hysteresis similar to that of normal ferroelectric [22]. Fig. 4(a) shows an example of the characteristic hysteresis loop in this temperature range. In the virgin state as indicated by the thick red line, it appears that there is no remnant polarization in the crystal at zero electric field. However, as increasing the electric field, we can see the gradual growth of polarization. This is a characteristic behavior of the polarization reversal (switching) in ferroelectric. When the applied field is larger than the coercive field $E_c$, ferroelectric domains are aligned along the direction of the electric field, leading to a stable macroscopic polarization in the crystal. This is evident from the fact that the remnant polarization is identical to the saturation polarization. These results clearly indicate that PMN is a ferroelectric rather than a dipole glass at low temperature.

There are many reports on the electric-field induced phase transition in PMN. On the basis of the change of dielectric constant under the application of a DC electric field, Colla et al.

Running Title

1  proposed an *E-T* phase diagram for PMN [8], which suggests a phase transition from glass
2  phase to ferroelectric phase at a critical field of $E_t$=1.5kV/cm in the temperature range of 160
3  K - 200 K. However, from our polarization results, we cannot find such a critical field except
4  the coercive field $E_c$. If we assume that $E_c$ is the critical field, then its value completely
5  disagrees with the reported value. We observed that $E_c$ increases rapidly with lowering the
6  temperature, for example, $E_c$ can reach a value of ~11 kV/cm at 180 K, which is about 7
7  times of $E_t$. Moreover, $E_c$ is strongly dependent on frequency (Fig.4b), having an exponential
8  form $(1/f \propto \exp(\alpha/E_c)$ (Fig.4c), resulting in an undefined critical field because of its
9  dependence on the measurement time-scale.

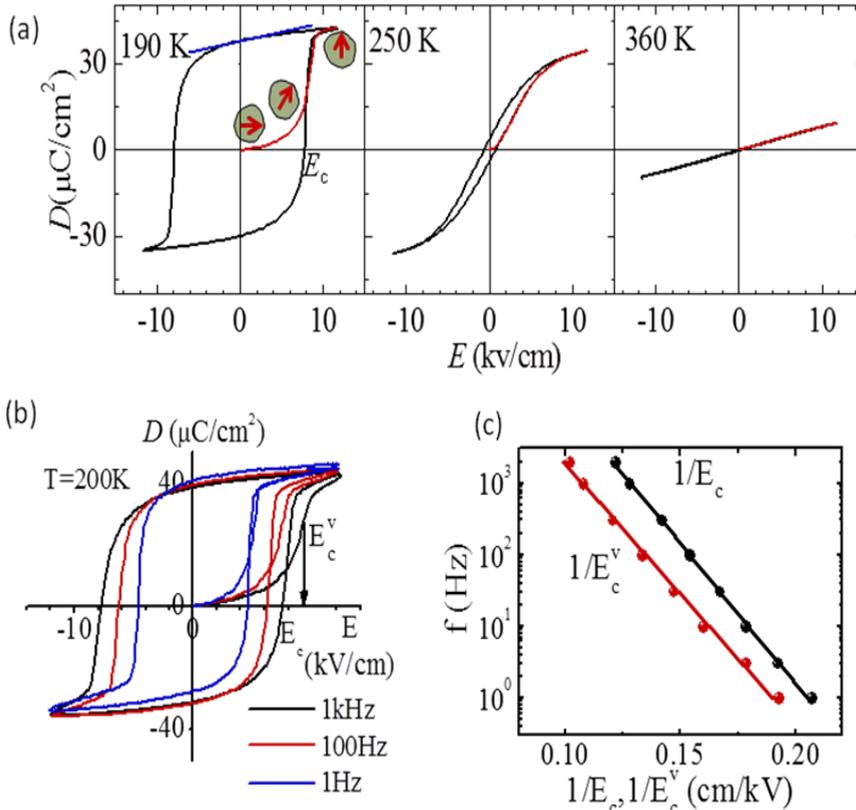

10
11  **Fig. 4.** (a) Polarization hysteresis in PMN crystal at 190, 250, 360K. (b)Frequency dependence
12  of polarization hysteresis. (c) Relationship between frequency and coercive field determined
13  from the peak of switching currents. Superscript $v$ denotes the value observed for the
14  virgin state.

15  As discussed in following, ferroelectric micro-domain and soft-mode behaviors have also
16  been observed at zero field in PMN in our measurements. Also, lowering of symmetry of
17  local structure at zero field was also revealed around 210 K by a NMR study [21]. All these
18  results direct to the fact of occurrence of a ferroelectric state at zero field in PMN crystal. We therefore consider that it is more rational to attribute $E_c$ to the coercive field required for the domain switching rather than the critical field for field-induced phase transition. In fact, the exponential relationship of coercive field with frequency is well-known as Merz's law ($f=1/\tau \propto \exp(-a/E_c)$, $\tau$=switching time, $a$=activation field) [23-28] in the normal ferroelectrics (for example, BaTiO$_3$,TGS) [23,24].   Using this relationship, the activation field is estimated to be 83.5 kV/cm for PMN crystal at 200 K (Fig. 4c), which is one order of magnitude greater than that of BaTiO$_3$ crystal.   This indicates that the domain switching in PMN become more and more difficult as lowering temperatures. For example, when an electric field of 1 kV/cm is applied to the crystal at 200 K, an unpractical time of $2.3\times10^{29}$ s ($\sim7.3\times10^{21}$ years) is required for the domain switching. Actually, it is impossible to observe the spontaneous polarization by this weak field within a limit time at this temperature.

The high-resolution data of the polarization obtained by a 14-bit oscilloscope allow us to calculate the linear and nonlinear dielectric susceptibilities (defined by the expansion $P=\varepsilon_0(\chi_1 E+\chi_3 E^3+\cdots)$) directly from the $D$-$E$ hysteresis by differentiating the polarization with respective to the electric field. The calculated results are summarized in Fig. 5 for various electric fields in the virgin state and the zero electric field after the polarization reversal. These results allow us to have a deep insight into the nature of abnormal dielectric behaviors and the phase transition in PMN relaxor. Fig. 5(a) shows the linear dielectric response for the virgin state in various electric fields. When comparing the response obtained at zero field with that obtained by LCR meter at the frequency corresponding to the sampling rate used in $D$-$E$ hysteresis measurements, one might find that they both behavior in the same way with the temperature. As mentioned above that the polarization response under the electric field in the virgin state is essentially due to the polarization reorientation, we therefore can reasonably attribute the dielectric anomaly usually observed in PMN relaxor to the polarization reorientation. This indicates that the reorientation of the PNRs dominates the huge dielectric response in PMN relaxor. Fig. 5 (a) also shows that the peak of dielectric response shifts to lower temperature at a higher electric field. This means that the activation field required for domain switching increases with lowering the temperature.

The nonlinear dielectric susceptibility $\varepsilon_{3v}$ and its scaled value $a_{3v}$ for the virgin state are given in Fig. 5(c) and (d), respectively. $\varepsilon_{3v}$ shows a broad peak around 255 K, which is in good agreement with those obtained by Levstik et al. by a lock-in to wave analyzer technique for various frequencies [9]. Levstik et al. attributed this behavior to the freezing of dipole glass at $T_f$=224 K in PMN. However, this picture is inconsistent with the results shown in the above polarization measurements, which indicates that PMN relaxor is ferroelectric but not glass at $T<T_f$. Such dipole glass picture is also excluded by the results of the scaled susceptibility $a_3$ shown in Fig. 5(d), and those reported in previous studies [29, 30]. We can see that there is no divergent behavior of $a_3$ in PMN relaxor. This result again suggests that there is no freezing of dipole glass in PMN as predicted for the dipole glass model. The observed anomaly of nonlinear dielectric susceptibility around 255 K is more consistent with the phase diagram of the random field model proposed by Fisch [13], in which a glass phase occurs between the paraelectric phase and the ferroelectric phase in relaxors. Such anomaly of nonlinear dielectric susceptibility may be a manifestation of spherical dipole glass with random interaction used in SRBRF model, and its nature requires further theoretical investigations.

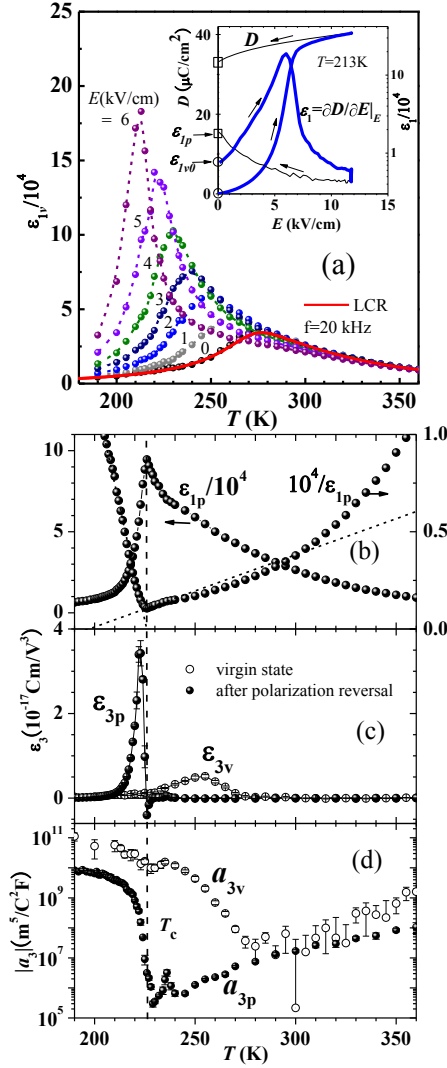

**Fig. 5.** Dielectric responses in PMN crystal. (a) Chang of linear dielectric responses ($\varepsilon_1 = \partial D / \partial E |_E$) with the electric field for the virgin state. Inset shows an example of the polarization and dielectric responses, in which the thick lines indicate the virgin state. Dielectric constant obtained by LCR impedance measurements at an ac level of 1 V/cm (red solid line) is also shown for comparison. (b) Linear dielectric response ($\varepsilon_{1p} = \partial D / \partial E |_{E=0}$) at zero field after the polarization reversal (indicated by the square in inset of (a)), and its inverse. (c) Nonlinear dielectric constant ($\varepsilon_3 = \partial^3 D / \partial^3 E |_{E=0}$) at zero field for the virgin state, and the state after the polarization reversal. (d) The corresponding $a_3 = \varepsilon_3 / \varepsilon_1^4 |_{E=0}$ for these two states.

In the random field model, the ferroelectric phase transition is suggested to be smeared due to the quenched random fields, but it may be visible if the random fields are overcome by an external electric field [12]. This ferroelectric phase transition has been convincingly shown by the sharp peak of the linear dielectric susceptibility $\varepsilon_{1P}$ (Fig. 5(b)) and the anomaly of its nonlinear components $\varepsilon_{3P}$ and $a_{3P}$ (solid circles in Fig. 5(c) and (d)) when ferroelectric domains are aligned by an external field. The sharp peak of linear susceptibility indicates that the ferroelectric phase transition occurs at $T=T_c$=225 K, around which Curie-Weise law was observed. The value of Curie constant is estimated to be $C$ = 2.05 *$10^5$ K, which is characteristic of that of the displacive-type phase transition, suggesting a soft mode-driven phase transition in this system. This conclusion is supported by the occurrence of soft–mode in the crystal observed by neutron [31] and Raman scattering measurements [32].

Here, we can see that there are two characteristic temperatures in relaxors: Burns temperature $T_{Burns}$ and ferroelectric phase transition temperature $T_c$. At $T_{Burns}$, local polarizations (PNRs) begin to occur. Before the ferroelectric transition occurs, PNRs are dynamic, and more importantly, interactions among them are random. Consequently, the existence of PNRs between $T_c$ and $T_{Burns}$ can be considered as precursor phenomenon in a phase transition [33]. Actually, such precursor phenomenon has also been observed in the normal ferroelectric BaTiO$_3$ in temperatures far above $T_c$ [34]. A difference between BaTiO$_3$ and PMN relaxor is that the temperate region of precursor existence in PMN is greatly lager than in BaTiO$_3$. To probe the precursor behavior, one has to consider the time scale used. For example, we can't detect a spontaneous polarization at ms scale by $D$-$E$ measurements for PMN at room temperature, but in the probe time scale of 2 ns, cold neutron high-flux backscattering spectrometer can detect PNRs up to ~400K [35] (Fig. 1b). In contrast, due to its short probe time scale (~6 ps) [4], thermal neutron scattering can probe PNRs up to 600 K, close to the $T_{Burns}$ determined by the optical measurements with the shortest probe time scale.

## 4. Soft mode behaviors in PMN relaxor

In the displacive-type ferroelectrics, soft-modes should occur in the lattice dynamics of the crystal. Actually, in a study of neutron inelastic scattering, a FE soft mode was revealed to recovers, i.e., becomes underdamped, below 220 K, and from there its energy squared $(\hbar\omega_s)^2$ increases linearly with decreasing $T$ as for normal FEs below $T_c$ (see also Fig. 6(e)) [31]. This has long been a puzzle for PMN relaxor: how can this be, since it has been thought that PMN remains cubic to at least 5 K [16]? However, such soft-mode behavior is exactly consistent with our results from polarization measurements shown in previous section, which show a ferroelectric phase transition at $T_c$~225 K. Our Raman scattering measurements also support the occurrence of FE soft mode in PMN relaxor [32].

In Raman scattering studies for relaxors, the multiple inhomogeneities due to the coexistence of different symmetry regions such as the PNR and COR has been a tremendous barrier to clarify the dynamical aspect of relaxor behavior in PMN. In particular, the intense temperature-independent peak at 45 cm$^{-1}$ (indicated by↓in Fig. 6(a)), which stems from the COR with $Fm\bar{3}m$ [32], always precludes a detailed investigation of low-wave number spectra of PMN crystal. Our angular dependence of the Raman spectra together with the results from the Raman tensor calculations clearly indicate that the strong $F_{2g}$ mode located

Running Title

1  at 45 cm⁻¹ can be eliminated by choosing the crossed nicols configuration with the
2  polarization direction of the incident laser along <110>c direction (see right panel in Fig. 6(c)).

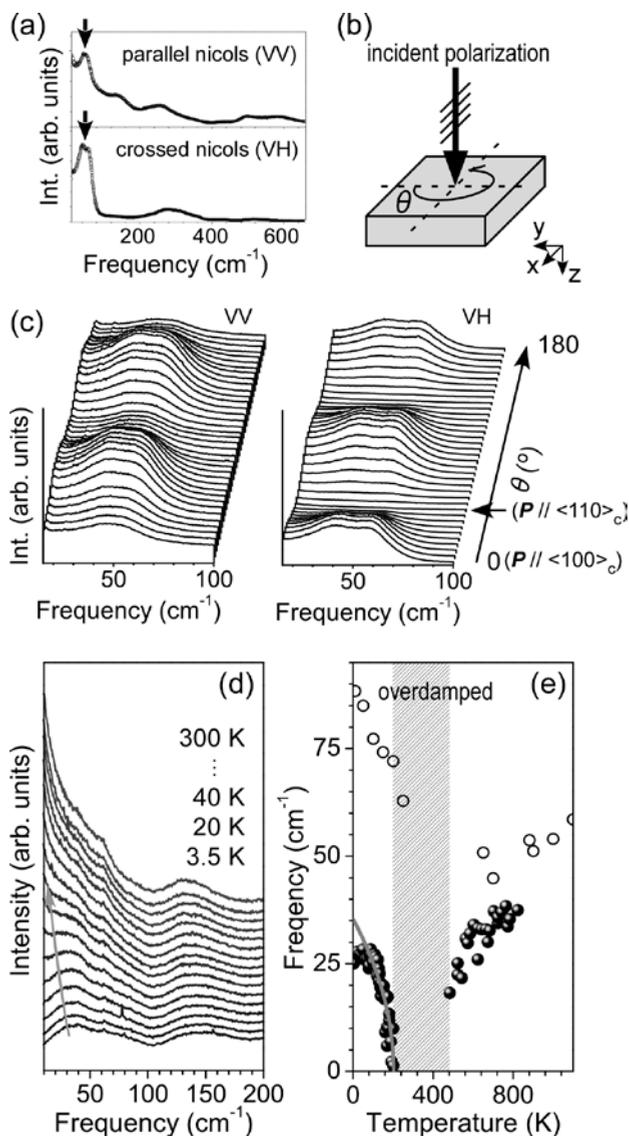

3
4  **Fig. 6.** (a) Room-temperature Raman spectra in PMN observed by the parallel (upper panel)
5  and crossed (bottom panel) nicols configurations, with the polarization direction of the
6  incident laser parallel to <100>c ($P$ // <100>c). (b) Scattering configurations used in
7  measurements on the angular dependence of the Raman spectra. (c) Angular dependence of
8  the low-wave number Raman spectra obtained at room temperature. (d) Temperature dependence of Raman spectra observed by the crossed nicols configuration with the polarization direction of the incident laser along <110>c direction (*P* // <110>$_c$). (e) Soft mode wave numbers obtained by Raman scattering (●) and neutron inelastic scattering by Wakimoto et al. (○) [31].

Such special configuration allows us to observe the other low-wave number modes easily. Fig. 6 (d) shows the spectra obtained by this configuration. At the lowest temperature, a well-defined mode can be seen from the spectrum, which softens as increasing the temperature, indicating the occurrence of FE soft mode in PMN relaxor. Due to the multiple inhomogeneities of the system, the shape of soft-mode of PMN relaxor is not as sharp as that observed in normal displacive-type ferroelectrics. However, we still can estimate its frequency reliably from the careful spectrum analysis. Its temperature dependence is shown in Fig. 6(e) (indicated by ●) in comparison with the results obtained by neutron inelastic scattering (○) [31].

We find that the soft-mode exhibits softening towards $T_c$ on heating and follows the conventional Curie–Weiss law (solid line) within a large temperature region. However, upon further heating, the soft mode becomes overdamped in a temperature region extending over ~200 K, which doesn't allow us to estimate the frequency of the mode. At temperatures above 480 K, the soft mode recovers the underdamped oscillation and hardens as the temperature   increases. These phenomena are very similar to those revealed by neutron inelastic scattering [31]. A major difference is that the wavenumber of the soft mode in the present study is significantly lower than that observed by the neutron inelastic scattering. This can be reasonably understood by the splitting of the soft mode due to the lowering of symmetry as demonstrated in the NMR study [21]. According to previous results, the local symmetry in the PNR changes from cubic to rhombohedral. Therefore, the soft mode can be assumed to split from the $F_{1u}$ mode to the $A_1$ and $E$ modes. Generally, the $A_1$ mode is higher in wave number than the $E$ mode due to the depolarization field effect.

In a short summary, we may say that the polarization in PMN is induced by the soft mode. This interpretation is essentially consistent with the results described in the polarization measurements, and the results obtained in previous neutron studies [36, 37], in which the crystallographic structure of PNR is attributed to the displacement pattern of the soft mode. The results of the Raman study also support that a ferroelectric state exists in PMN even at the zero-bias field.

## 5. Ferroelectric domain structures observed by TEM

In order to understand the microstructures of COR and PNR together with the domain structures and its evolution with temperature in the ferroelectric phase of PMN relaxor, we have carried out a detailed TEM observation. The typical results are summarized in Fig. 7. As shown in Fig. 7 (a″)-(c″), COR was found to be spherical shape and has size less than 5 nm. It is very stable and remains unchanged within the temperature range of 130-675K. In the TEM observation, large amount of CORs were found to distribute in the PMN crystal. In a previous HRTEM study, its volume fraction has been estimated to be ~1/3 of the crystal [18]. CORs are thus considered to be the intense sources of the strong random fields.

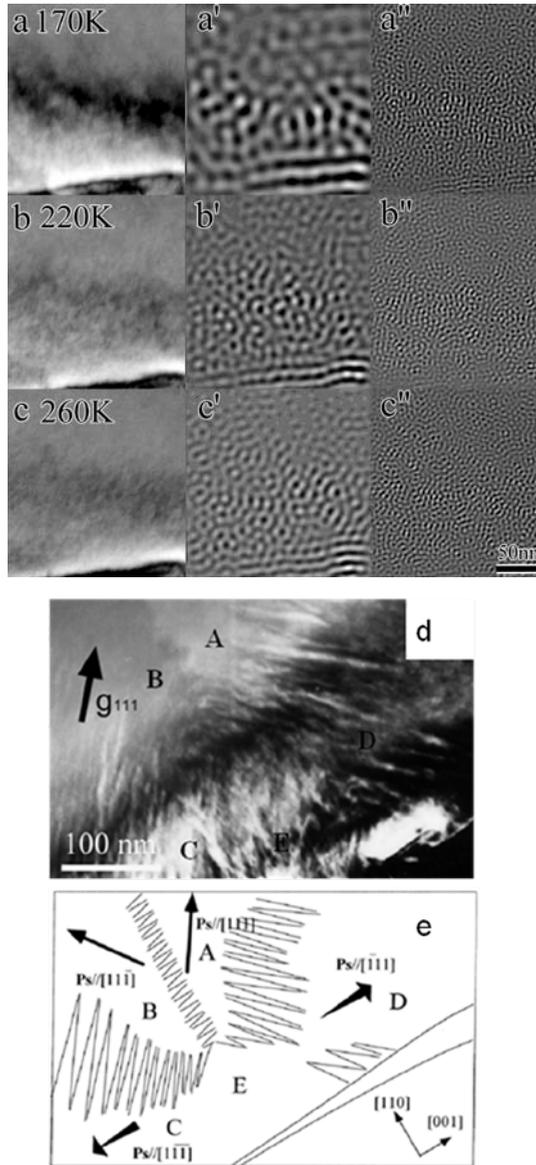

**Fig. 7.** (a)-(c) Temperature variation of TEM images observed for PMN relaxor. (a′)-(c′) & (a″)-(c″) show images of PNR and COR derived from (a)-(c). (d) Micrometric domain structure observed in the ferroelectric phase at 130K. (e) Schematic domain patterns shown in (d). Arrows and lines indicate the polarization directions and domain boundaries, respectively.

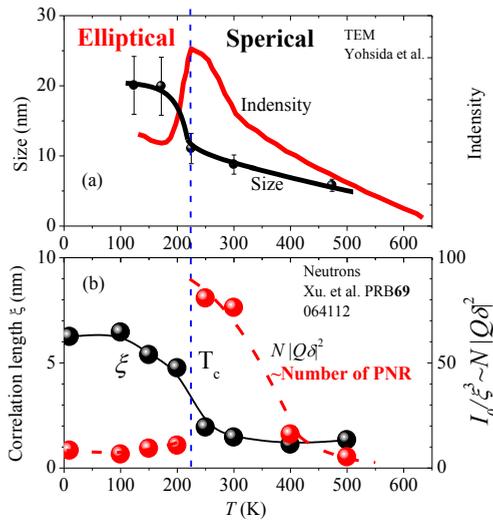

**Fig. 8.** Temperature variation of PNR observed by (a) TEM [38] and (b) Neutron scattering [4]. In (b), $\xi$ is the correlation length, and $I_0$ is the integrated diffuse scattering intensity and can be written to $N\xi^3|Q\delta|^2$, where $N$ is the total number of PNR, $\delta$ is the average displacement of atoms within the PNR, and $Q$ is the wave vector.

In contrast to COR, PNRs exhibit remarkable change with temperature. As shown in Fig. 7(a′)-(c′) and Fig. 8, PNRs with size of several nm were found to occur in the crystal for $T<T_{Burns}$. These spherical PNRs show continuous growth as lowering the temperature. However, PNRs change from spherical shape to elliptical shape around $T_c$. Associating with the change in shape, its intensity was also found to drop rapidly. These results are consistent with those derived from neutron scattering (Fig. 8(b)) [4]. Neutron scattering study by Xu et al.[4] shows that the ''correlation length'' $\xi$, which is a direct measure of the length scale of the PNR, increases on cooling and change remarkably around $T_c$. At the same time, the numbers of PNR increases on cooling from high temperatures and then drops dramatically at around $T_c$, remaining roughly constant below $T_c$.

Associating with the change with PNR, micrometric ferroelectric domains were found to occur for $T<T_c$. It is because of growing into macroscopic domain that the numbers of PNR drop sharply in the ferroelectric phase. Figs. 7(d) and (e) show the structure of a FE micrometric domain in the FE phase of 130 K and its schematic patterns, respectively. The micrometric domains are formed in the crystal with the spontaneous $P_s$ along the <111>$_c$ direction. In comparison with the domain structure of normal ferroelectric such as $BaTiO_3$, domain size is relatively small and the domain boundaries blur in PMN relaxor.

The occurrence of micrometric domains, the soft-mode observed by Raman scattering, and the macroscopic polarization all direct to the same conclusion: PMN is essentially ferroelectric but not dipole-glass at $T<T_c$ although it exhibits some unique characteristic properties including broadening soft-mode, smearing domain wall, and very large activation field required for domain switching. Our TEM measurements clearly indicate that interactions among PNRs for $T<T_c$ are not random, but cooperative, which is different from

1  the picture expected in (SRBRF) model [10]. It is due to such **non-**random interaction, PNRs
2  team up together to form micrometric domain in the FE phase. Thus, our TEM observations
3  support the random field model suggested by Fisch [13]. It should be emphasized that
4  PNRs cannot merge together completely due to the blocking of the intense CORs in the
5  crystal (Fig. 10).

6  Here, we made a discussion on the volume fraction of PNRs in PMN crystal. Neutron
7  scattering technique has been used to estimate the volume fraction of PNRs. Fig. 9 replots
8  two results reported by Jeong et al.[37] and Uesu et al.[39], respectively. Both studies
9  indicate that PNRs occupy a volume fraction > 25% at the lowest measurement temperature.
10 This volume fraction is larger than the threshold of 22% to form a percolated ferroelectric
11 state with an ellipsoidal-shape [40], supporting again the picture of a ferroelectric state in
12 PMN relaxor.

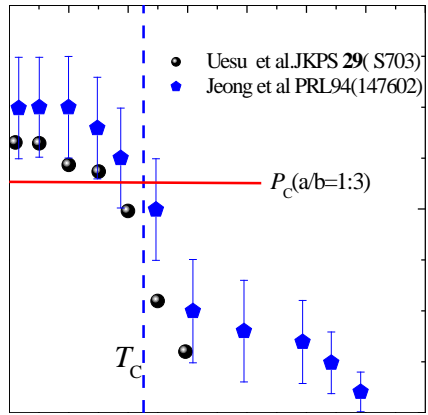

13
14 **Fig. 9.** Volume fraction of PNR estimated from neutron scattering measurements [37,39].
15 Solid line denotes the threshold of percolation for elliptical shape [40].

16 ## 6. A physics picture of relaxor

17 In summary, we propose a physics picture for relaxors. Figure 10 schematically shows a
18 model of structure evolution in PMN relaxor. Since COR has been observed at $T > T_{Burns}$, it
19 can be considered that there is a coexistence of paraelectric phase of COR with $Fm\bar{3}m$
20 symmetry and paraelectric $Pm\bar{3}m$ phase in this high temperature. Upon cooling, spherical
21 PNRs occur from paraelectric $Pm\bar{3}m$ phase for $T < T_{Burns}$. Both number and size of PNR
22 increase as lowering the temperature. Around room temperature, PNRs grow to a size of
23 about 10 nm. For $T < T_c$(~225K), neighboring PNRs merge together to form elliptical shape
24 with anisotropy, associating with the reduction of its number. Due to blocking by the intense CORs, individual PNRs merely grow to a size of about 20 nm in the low temperature. However, PNRs with elliptical shape tend to team up together to form larger domain in the ferroelectric state. Such multi-scale inhomogeneity of domain structure provides a key point to understand the huge electromechanical coupling effects in relaxors and piezoelectrics with morphotropic phase boundary (MPB). This also gives idea how to design new material having domain structure to enable large elastic deformation by an electric or magnetic field [41].

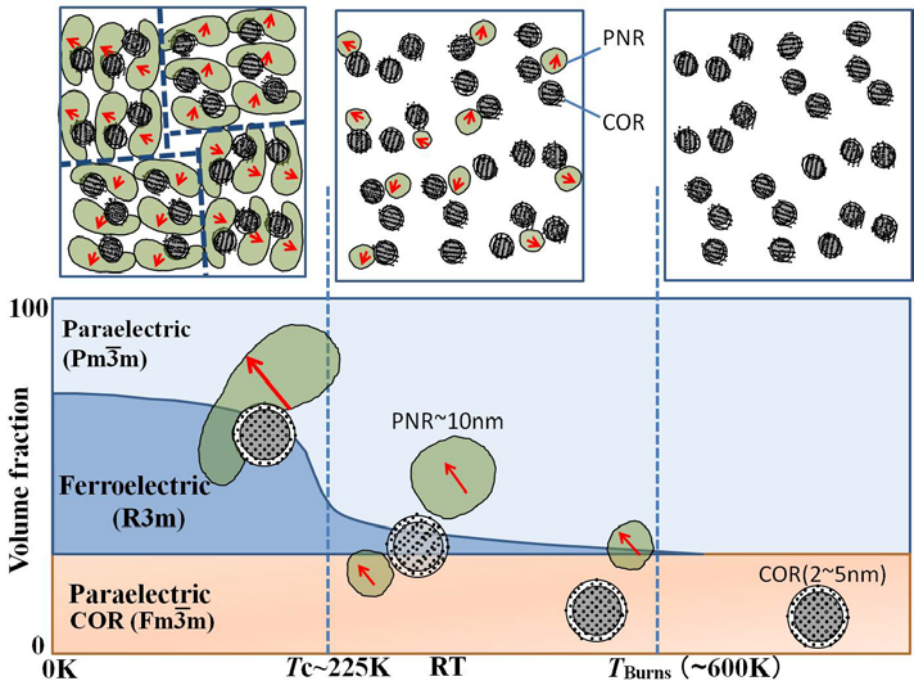

**Fig. 10.** A physics picture of structure evolution in PMN relaxor.

## 7. Acknowledgements

We thank Mr. M. Yoshida, Prof. N. Yamamoto and Prof. Shin-ya Koshihara of Tokyo Institute of Technology for their contributions in this work.